# SUMMARY OF WORKING GROUP ON SINGLE BEAM HIGH LUMINOSITY ISSUES

S. Guiducci, LNF, Frascati, Italy, C. Zhang, IHEP, Beijing, CHINA

*Abstract*

The aim of the Working Group on Single Beam Behaviour was to concentrate on the items limiting the achievement of high luminosity. Some are related to high current and short bunch distance, as electron cloud instability (ECI), the others to the lattice design, as Dynamic Aperture (DA), wigglers, Interaction Region (IR) design, lifetime and background. These arguments have been discussed to explore the feasibility of a very high luminosity Φ-factory.

## INTRODUCTION

Effect of wigglers on the beam and electron cloud instability, which affect the possibility of achieving high luminosity, was discussed in two dedicated talks. A general presentation on the Beijing τ-charm project, the BEPCII, has reviewed all the accelerator physics issues relevant to a low energy collider. Other talks have been focused on the study of an ultra high luminosity Φ-factory. For this design a new lattice, was presented and the problems related to the IR design, lifetime and background were discussed.

## WIGGLERS

To increase the beam-beam limit threshold it is necessary to have strong radiation damping. Wiggler magnets are used to increase radiation damping at low energy machines as, for example, in DAΦNE. A large number of superconducting wigglers (B = 2.1T, length 18.2 m) will be used in CESR-c to operate the collider at a lower energy as a τ-factory [1]. In order to avoid the harmful effects of nonlinearity in the magnetic field special care has been adopted in the following operations:

    Magnetic design and field calculation
    Magnetic measurements
    Simulations
    Beam based characterization.

Two versions of wiggler magnets with symmetric (7 poles) and asymmetric (8 poles) structure have been developed, built and tested. Magnetic field measurement revealed that the magnets with asymmetric structure have significantly less variation of integrated magnetic field properties with excitation.

Beam based characterization has been done by measuring the vertical beam size in the tune plane (see Fig. 1), showing the coupling resonances, and by measuring the tunes as a function of a displaced orbit. Beam based characterization of the wiggler magnets confirmed model calculation and results of magnetic field measurement.

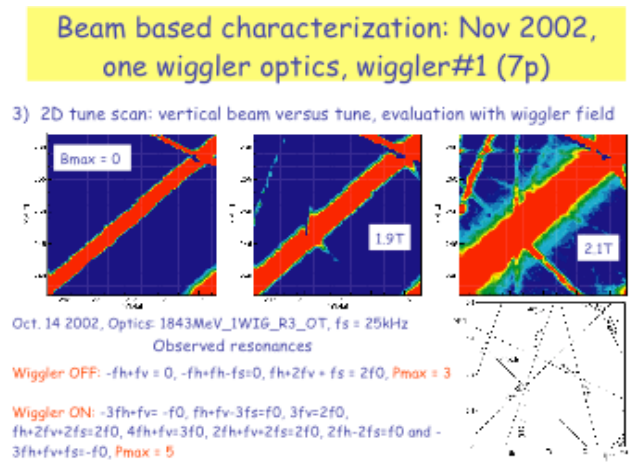

Fig. 1: Vertical beam size vs. tune at various wiggler fields.

## ELECTRON CLOUD INSTABILITY

A general description of the problem and the status of simulations and measurements [2] have been presented. In particular the work done at CERN and the simulation code developed there have been described. This instability is an important luminosity limitation for the B-factories. Up to now currents as large as 2A have been stored in DAΦNE and no evidence of ECI has been observed. In order to make predictions for future machines it is important to find an agreement between measurements and calculations. Study is needed on simulations to improve the model used, but also measurements on the electron yield in the vacuum chamber are very important as input to simulation codes. At LNF a solid state physics laboratory to make these measurements is in preparation. DAΦNE is a good benchmark for simulation codes because they foresee a large effect that has not been observed. Some effects that might explain the discrepancy, like the special shape of the vacuum chamber (with antechamber) and the fact that magnetic elements are very close to each other, have to be included in the code.

A new code, based on Ohmi's model, has been developed to evaluate this effect for BEPCII. Fig. 2 shows how the EC density is reduced for an elliptical pipe with antechamber. To guarantee the beam performance against ECI, precaution methods, successfully adopted in PEP-II and KEKB, have been foreseen:

Antechamber
TiN coating of the inner surface
Solenoid winding (as backup)
Clearing electrodes (R&D)

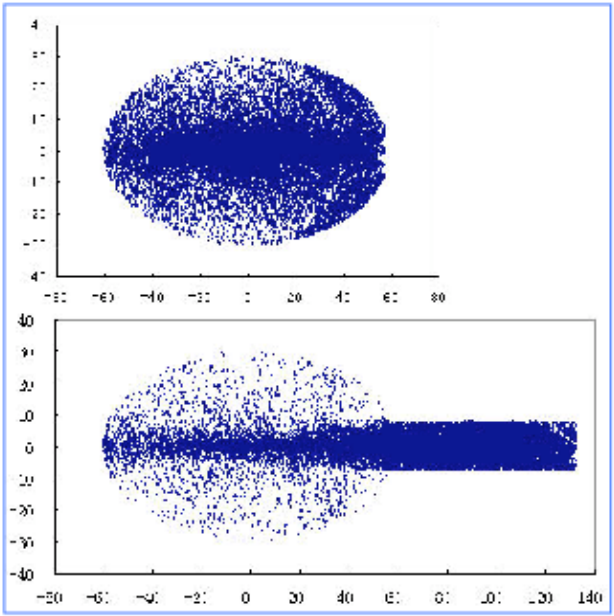

Fig.2: EC density with and without antechamber.

## ACCELERATOR PHYSICS ISSUES IN BEPCII

BEPC-II is the upgrade of the present BEPC to a double ring collider with a luminosity increase by a factor 100 at 1.89 GeV. It has to be operated also as synchrotron radiation source, at 2.5 GeV, using the existing beam lines. The main machine parameters are listed in Table 1. A very general review of the accelerator physics issues of the BEPCII has been given [3] including lattice, dynamic aperture, lifetime, ion trapping, ECI, impedance budget and instabilities, beam-beam effects.

Table 1- BEPC-II parameters

| Energy range | 1 ~ 2 GeV |
|---|---|
| Luminosity @1.89 GeV | $10^{33}$ cm$^{-2}$ s$^{-1}$ |
| Circumference | 238 m |
| Natural emittance | 0.14 μ rad |
| $\beta_x^*$, $\beta_y^*$ | 1.0, .015 m |
| Bunch length length | 0.015 m |
| Crossing angle | ± 11mrad |
| Bunch distance | 2.4 m |
| Current @1.89 GeV | 0.91 A |

A summary of the single beam effects study is given below:
- With the present impedance budget a bunch length less than $\beta_y^*$ is estimated
- Coupled bunch instabilities and fast beam ion instability in e$^-$ ring can be damped with a feedback system.
- A gap in the bunch train is needed to avoid ion trapping.
- Antechamber (TiN coated) is adopted to reduce EC; further study is underway.
- Single beam lifetime is about 3.1 hours; with top-off injection the average luminosity should be larger than $6\times10^{32}$ cm$^{-2}$ s$^{-1}$.

Beam-beam simulations have been done to choose the working point and study the effect of crossing angle. According to simulations the luminosity is reduced by a factor .8, due to hourglass and crossing angle, and the design value $\xi_y$ ~0.04 is reachable. Some further simulations should be done including coherent beam-beam effects by strong-strong simulations.

## ULTRA HIGH LUMINOSITY Φ-FACTORY

In a low energy collider as a Φ-factory, increasing currents and number of bunches is not enough to reach ultra high luminosity. The beams have to be squeezed at the Interaction Point (IP) as much as possible and the beam-beam tune shift has to be increased. Hence new ideas are needed.

At present there are two catalogues of approach for high luminosity : one is the methods used at present factories (two rings, crossing angle, flat beam, many bunches and very high current) and other one is proposed approach, such as the round beam scheme in Novosibirsk for VEPP2000 [4]. A lattice design based on the first approach has been presented in this session [5]. For this scheme considerations on the IR design [6], the expected beam lifetime [7], and the machine background in the detector have also been presented [8].

### Lattice

To achieve ultra short bunches at the IP a new idea to use strong longitudinal focussing in order to have a bunch length variation along the ring [9] has been proposed. With this scheme the bunch can be very short only at the IP and longer elsewhere in the ring, where the vacuum chamber impedance is critical and produces lengthening and instabilities.

The plot of RF voltage, bunch length at IP ($\sigma_{IP}$) and at RF cavity ($\sigma_{cav}$) versus the longitudinal phase advance is shown in Fig. 3. The bunch lengths $\sigma_{IP}$ and $\sigma_{cav}$ are equal for small phase advance; when the longitudinal phase advance $\mu_l$ increases, the bunch length at the IP decreases while that at the cavity $\sigma_{cav}$ increases. Very high RF voltage and large and negative [10] momentum compaction are needed to achieve very small bunch length at IP.

A lattice suitable for the strong longitudinal focusing has been presented. In order to get large (negative) momentum compaction, keeping the dispersion function small, the cell is made of alternating positive and negative dipoles (see Fig. 4). In this way a very high radiation damping (a factor 3 larger than present DAΦNE value) is obtained without using wigglers, which can have harmful effect on the beams due to field nonlinearity.

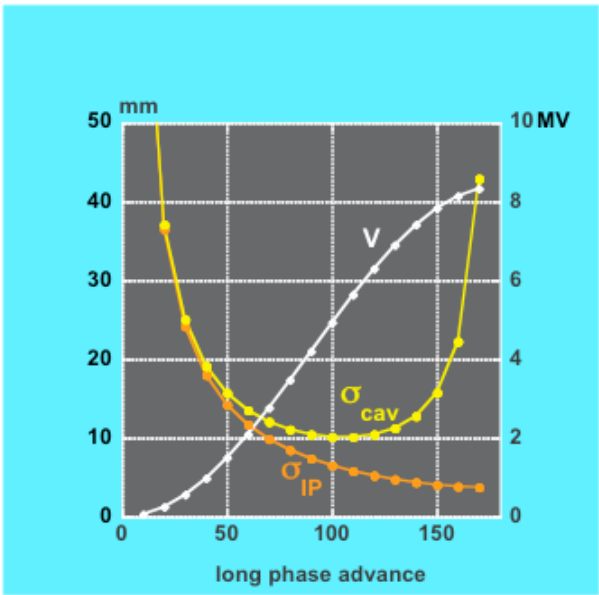

Fig.3: RF voltage and bunch length vs. longitudinal phase advance.

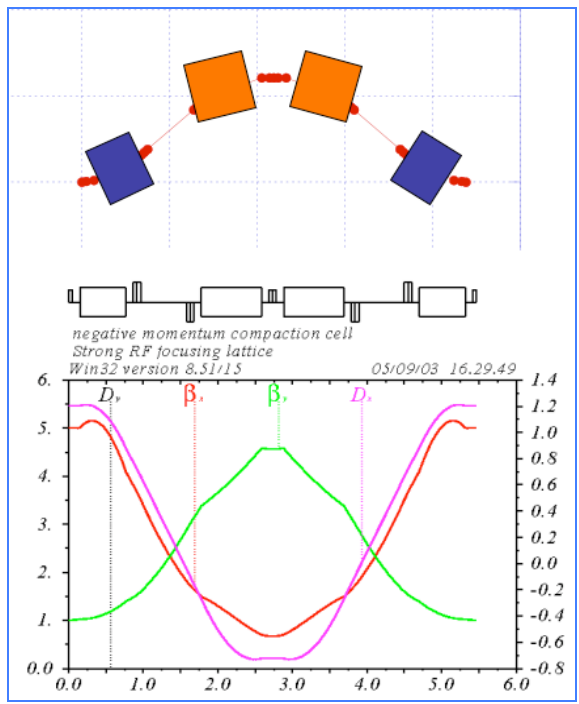

Fig. 4: Lattice cell

The two rings layout is similar to the present DAΦNE one and fits in the same hall. There is only one IR; the opposite straight section, where the beams are separated, can be used for injection, RF and diagnostics.

DA is a critical item for this machine. The lattice cell is very good for chromaticity correcting sextupoles because they are placed where the dispersion and the separation between $\beta_x$ and $\beta_y$ is maximum. The resulting DA, without synchrotron oscillations, is very large; nevertheless when synchrotron oscillations are taken into account the DA is strongly reduced, showing a strong dependence on the RF voltage and synchrotron tune due to resonances in 3-dimensional space. This dependence has to be further explored for the strong RF focusing parameters to see if a proper choice of the working point can improve the DA.

*IR design*

The design of an IR with a very small vertical beta is a challenging task and is a common Machine & Detector business. The following list of constraints and design requirements has been presented; where (D) means detector and (M) machine.

Maximum detector solid angle, try to keep accelerator components far enough from the IP (D)
Large high-field detector solenoid (D)
Push first quadrupole close to IP, to minimize IP spot size (M)
Horizontal crossing angle (M) (DAΦNE experience)
Small quadrupoles, inside detector field (M,D)
Coupling correction (M) (DAΦNE experience)
Adequate shielding from background (M,D)
Ultra-vacuum (M,D)
Impedance budget (M)
Thin beam pipe (D)
"Instrumented" IR (D)

The choice of crossing angle has been discussed taking into account the effects of parasitic crossings, the reduction of tune shifts and luminosity and the constraints due to the IR geometry and aperture requirements.

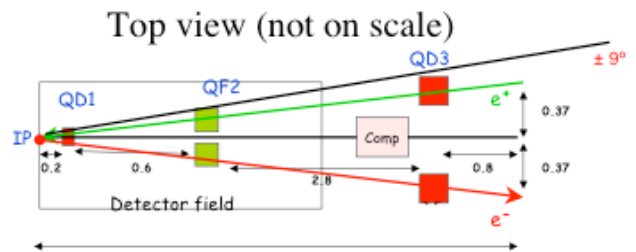

Fig. 5: Preliminary layout of half IR

A preliminary design, with permanent magnet quadrupoles, has been presented (see Fig. 5). Machine elements are inside a ±9° cone with ±30 mrad crossing angle. The vertical beta at the IP can be varied between 1.5 and 5 mm.

As a conclusion the following list of items, which need to be addressed in the future, has been given:

Technical design
Engineering studies of permanent quads
Chromaticity correction study
Coupling correction scheme
Background evaluation
Beam pipe design
Vacuum design
Impedance budget
Trapped HOM study
Temperature control

*Lifetime*

At the Φ energy, beam lifetime is mainly limited by Touschek scattering. Touschek lifetime is inversely proportional to the bunch density and therefore it decreases essentially as the inverse of luminosity. It is also nearly proportional to the square of the energy acceptance of the ring. Therefore in the design of such a machine it is very important to increase as much as possible the energy acceptance, which is limited by vacuum chamber aperture, DA and RF acceptance. In the strong focusing case the RF parameters, and therefore the energy acceptance, are determined by the requirements on the bunch length.

Expected single beam lifetimes have been calculated [10] for the proposed lattice in different configurations. In each case lifetime is of the order of ten minutes at a luminosity of $10^{34}$ cm$^{-2}$ s$^{-1}$. Thus continuous injection, with detector on, is required. Moreover a procedure to optimize the machine with a rapidly decreasing current must be studied.

*Detector background*

A preliminary simulation of the trajectories of the Touschek scattered particles [11] for the proposed lattice has been shown (see Fig. 6).

Sensitivity to different parameters has been presented. The background in the detector poses some constraints on the IR design, in particular the aperture of the IR vacuum chamber has to be increased in the critical positions where particles get lost.

A system of scrapers (very effective in DAΦNE) has to be studied to reduce the background. In particular the positions of the scrapers have to be optimized by tracking. Moreover, as was done for DAΦNE, the particles lost in the IR have to be tracked inside the detector, by a Monte Carlo simulation, in order to evaluate the effective background and to study the design of tungsten masks to reduce it.

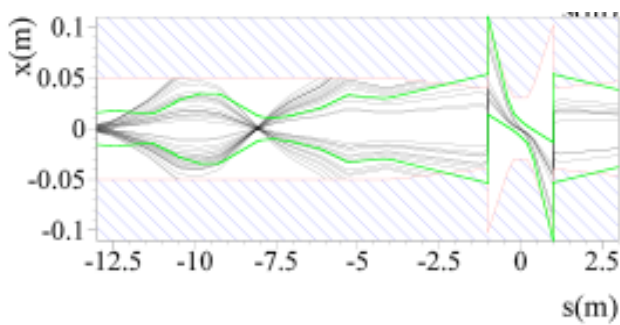

Fig. 6: trajectories of Touschek scattered particles near the IR (the IP is at zero)

# CONCLUSIONS

Aspects of single beam behaviour critical for the achievement of very high luminosity have been presented. The feasibility of an ultra high luminosity Φ-factory has been discussed. A preliminary design, based on dipoles with alternating polarity and strong longitudinal focusing, has been presented. This scheme is promising but more work is still needed. As a result of the discussion, a list of the items to be still addressed is given below:

Optimize longitudinal parameters and define the RF system.
More simulations:
  - Dynamic aperture with Synchrotron oscillation
  - Magnetic errors and fringing fields
  - Longitudinal dynamics
  - Impedance budget
  - Beam-beam effect
Test of Negative $\alpha_c$ at DAΦNE (done at KEK B-factory [11])
Test of longitudinal strong focusing
Final IR design
Lifetime: check the agreement between simulations and measurements at DAΦNE
Instability and feedbacks
Dipole design
Design of superconducting IR quadrupoles.